\documentclass[11pt]{article}

\makeatletter
\def\ps@top{\let\@mkboth\@gobbletwo
     \def\@oddhead{\rm\hfil\thepage\hfil}\def\@oddfoot{}
     \def\@evenhead{}\let\@evenfoot\@oddfoot}
\def\@bibsetup{\itemindent=-\leftmargin}
\def\@citesep{; }
\def\@cite#1#2{({#1\if@tempswa , #2\fi})}
\def\@biblabel#1{\hfill}
\def\thebibliography#1{\section*{References\markboth
 {REFERENCES}{REFERENCES}}\list
 {[\arabic{enumi}]}{\settowidth\labelwidth{[#1]}\leftmargin\labelwidth
 \advance\leftmargin\labelsep
 \usecounter{enumi}\@bibsetup}
 \def\newblock{\hskip .11em plus .33em minus -.07em}
 \sloppy
 \sfcode`\.=1000\relax}
\renewcommand{\section}{\@startsection {section}{1}{\z@}{-3.5ex plus -1ex minus 
    -.2ex}{2.3ex plus .2ex}{\centering\large\bf}}
\renewcommand{\subsection}{\@startsection{subsection}{2}{\z@}{-3.25ex plus
    -1ex minus -.2ex}{1.5ex plus .2ex}{\centering\bf}}
\makeatother
\begin{document}
\begin{center}{\large\bf Atomic Oxygen in the Comae of Comets} \\ [15pt]
Anita L. Cochran \\
University of Texas at Austin \\
McDonald Observatory \\
1 University Station, C1402 \\
Austin, TX 78712-0259 \\ [5pt]
Accepted for Icarus \\ [1in]
\end{center}

\begin{center}{\bf Abstract}\end{center}
We report on the detection of atomic oxygen lines in the spectra of 8
comets.  These forbidden lines are a result of the photodissociation
of the parent oxygen-bearing species directly into an excited state.  We
used high resolution spectra obtained at the McDonald Observatory 2.7m
telescope to resolve the cometary oxygen lines from the telluric oxygen
lines and from other cometary emissions.  We find that the relative
intensities of the two red lines (6300.304 and 6363.776~\AA)
are consistent with theory.  The green line (5577.339~\AA)
has an intensity which is about 10\% of the sum of the intensities of
the two red lines. 
We show that collisional quenching may be important in the inner coma.
If we assume the relative excitation rates of potential
parents which have appeared in the literature, then H$_{2}$O would be the
parent of the cometary green oxygen line.  However, those rates have been
questioned.  We measured the width of the three oxygen lines and find that
the green line is wider than either of the two red lines. 
The finding of a wider line could imply a different parent for the green
and red lines.  However, the constancy of the green to red line flux ratio
suggests the parent is the same for these lines but that the exciting photons
have different energies.


\newpage
\section{Introduction}
Oxygen is important for the chemistry of many of the current Solar System
bodies since it readily bonds with many other atoms. 
There are three prominent atomic oxygen lines in
the optical region of the spectrum.
The transition from the first excited state of
the singlet branch, $^1$D, to the ground state, $^3$P, produces
a doublet in the red at
6300.304~\AA\ and 6363.776~\AA.  The second excited singlet state, $^1$S,
decays to the $^1$D state 90--95\% of the time with a transition
at 5577.339~\AA; the remaining 5-10\% of the time, the $^1$S state decays
directly to the ground state ($^1$S -- $^3$P)
through two UV lines at 2958.365 and 2972.288~\AA\ (Slanger {\it et al.} 2006a).
\nocite{slangbranch06}
These are all forbidden lines.
The green line and the red doublet are observed in the Earth's atmosphere
and are affected by auroral activity. 
The green line has been observed in spectra of Venus and appears to be
quite variable there (Slanger {\it et al.} 2006b). \nocite{slangvenus06}
It is assumed that these lines will also be present in Mars' atmosphere,
but only the UV line at 2972.288~\AA\ has been detected definitively
(Barth {\it et al.} 1971; Leblanc {\it et al.} 2006).
\nocite{barthetal71,leblanc06}

The three forbidden lines are also seen in the spectra of comets.
Photodissociation of oxygen-bearing parent species produces
oxygen atoms in the ground $^3$P state or directly into the
excited $^1$D or $^1$S states, depending on the parent
molecule and the nature of the solar photodissociation.
The red doublet is substantially stronger than the green line
and the red lines are generally much stronger than the telluric red lines.
The cometary green line is generally a small fraction of the intensity
of the telluric green line.

To observe the oxygen in the spectra of comets, it is desirable to observe 
with moderately high spectral resolution in order to resolve the 
Doppler-shifted cometary lines from the telluric lines.   The red doublet
occurs in a part of the cometary spectrum which is relatively uncrowded,
so contamination from other cometary species is not an issue at
high spectral resolution. The spectral resolution allows for the
separation of the cometary and telluric lines. 
The situation for the green line is quite
different.  In addition to the telluric green line which must be
separated, the cometary green line sits in the middle of the C$_{2}$ (1,2)
P-branch; the P$_1$(27) and P$_2$(26) lines flank the oxygen
line.  The C$_{2}$ spectrum is quite dense.  Thus, for many cometary
spectra, it is impossible to derive the intensity of the green 
line.  However, for some comets, the oxygen line is stronger than
the C$_{2}$ bands and can be successfully deblended from the
cometary C$_{2}$ and the telluric oxygen if the spectral resolution 
is sufficient.  Some comets have been observed to have almost
no C$_{2}$, making detection of the green line simple (Cochran and
Cochran 2001). \nocite{coco01}

The green line and the red doublet are not produced by fluorescence
or dissociative recombination (Festou and Feldman 1981) \nocite{fefe81}
but instead arise from atoms produced directly in the $^1$S or $^1$D states
(i.e.  prompt emissions).
The lifetimes of the upper states of the oxygen lines are short,
with the $^1$D state lasting
about $\sim110$\,s and the $^1$S state lasting $\sim1$\,s.
Indeed, the lifetime of the $^1$S state is so short that its
spatial distribution traces
the parent's distribution directly, since the oxygen atom will not have
time to travel far from its direct parent prior to its decay
to the ground state.

\begin{figure}
\vspace{4in}
\includegraphics{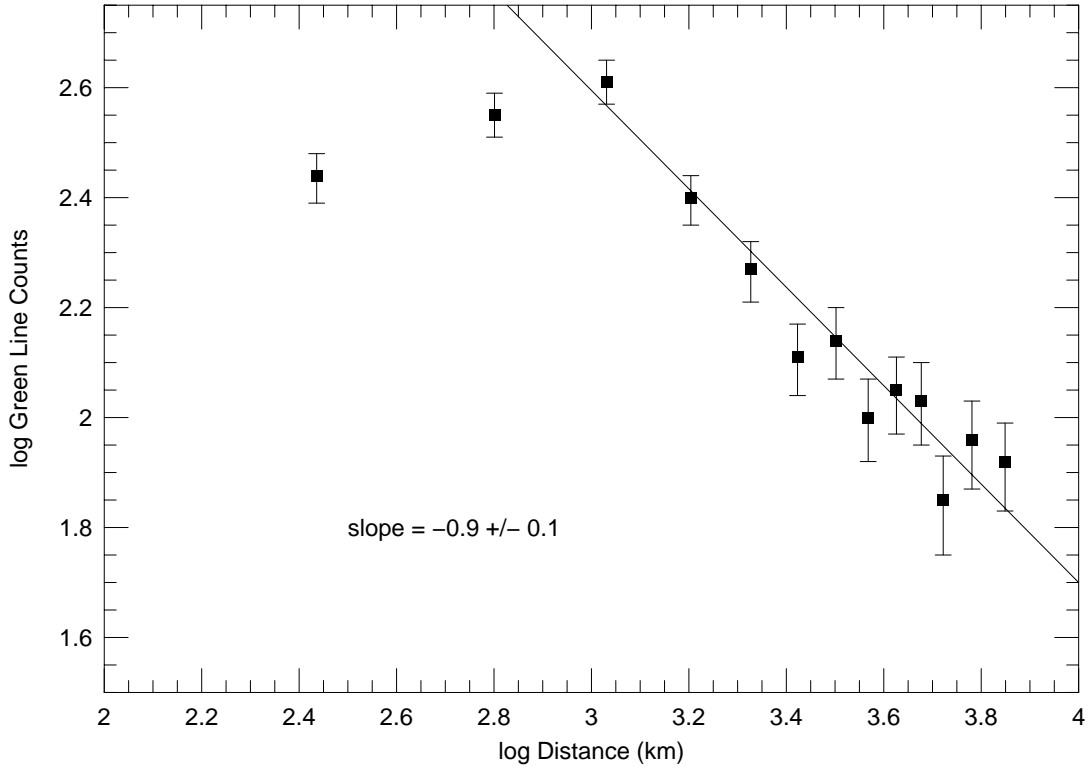}
\caption[fig1]{The intensity of the green oxygen
line at 5577.339~\AA\ was measured immediately after the impact
of comet 9P/Tempel~1 and the Deep Impact spacecraft using the
HIRES instrument at Keck I.  The green line initially increased
in intensity and then decayed as it flowed outwards.  This figure
shows the intensity of the line as a function of cometocentric
distance, $\rho$.  The line decayed as $1/\rho$, the signature
of a parent.  Since the lifetime of the $^1$S state is $\sim1$\,s at
1\,{\sc {\sc au}}, this points to a single-step process from parent to
O ($^1$S) state.}\label{tempel1}
\end{figure}

The atomic oxygen observed in the spectra of cometary comae did not
sublime from pure oxygen ice. Instead, oxygen is stored in the nucleus
as an oxygen-bearing ice.  Jackson and
Cochran (2007) \nocite{jaco08} observed the green line in the spectrum
of comet Tempel 1 following the impact of the Deep Impact spacecraft
with the comet.  They saw the line increase in brightness quickly
after the impact and then decay.  Figure~\ref{tempel1} shows the
intensity of the green line after the impact as a function of the
cometocentric distance. 
The initial increase shows the impulsive nature of the impact.
The decay started at $\sim2000$s after the impact.  After that time,
the green line flux decayed as 1/$\rho$,
where $\rho$ is the projected cometocentric distance.  
This behavior is characteristic of prompt emission produced by a 
single-step dissociation.  Since, with its short lifetime, the 
emission is tracing the direct parent, the behavior shown in
Fig.~\ref{tempel1} is indicative that the $^1$S state is a daughter, and
not a granddaughter, product of some species.
But what is this parent? 
Complex oxygen-bearing species (e.g. H$_2$CO or HCOOH) can
not be the parent, as they can only produce
oxygen via a two-step process.  In addition, Jackson and
Cochran showed that the green line flux decayed much
faster than the OH flux, indicating that the oxygen line did not
come primarily from OH.  Thus, the green line is
likely to be directly produced from H$_{2}$O, CO$_2$, or CO. 
It is generally assumed that the red lines come from the
dissociation of H$_{2}$O as this is consistent
with the distribution of  O ($^1$D) in the inner
coma of a comet (Combi and McCrosky 1991; Fink and Johnson 1984;
Magee-Sauer {\it et al.} 1988).
\nocite{comc91,maetal88,fijo84}
We have used high spectral resolution observations of 8 comets to
detect the green line and the red doublet of oxygen in order
to try to understand the production of the atomic oxygen in cometary
spectra.

\section{Observations}

The spectra of the comets were obtained with the 2DCoude instrument
at the 2.7m Harlan J. Smith telescope of McDonald Observatory.
This is a cross dispersed echelle spectrograph (Tull {\it et al.} 1995) 
which can be operated in two resolution modes.  The lower
resolving power has R=$\lambda / \Delta \lambda$ = 60,000.  In this
mode, we obtain complete wavelength coverage from 3700~\AA\ to
5700~\AA\ and continued coverage from 5700~\AA\ to 10,200~\AA\ with increasing
inter-order gaps.  Thus, the green and the red lines are observed
simultaneously.
The other mode of this instrument has R=200,000.
In this mode, the wavelength coverage is much smaller (1/16th of
the R=60,000 mode) but individual lines may be more easily resolved. 

\begin{figure}
\vspace{7in}
\includegraphics{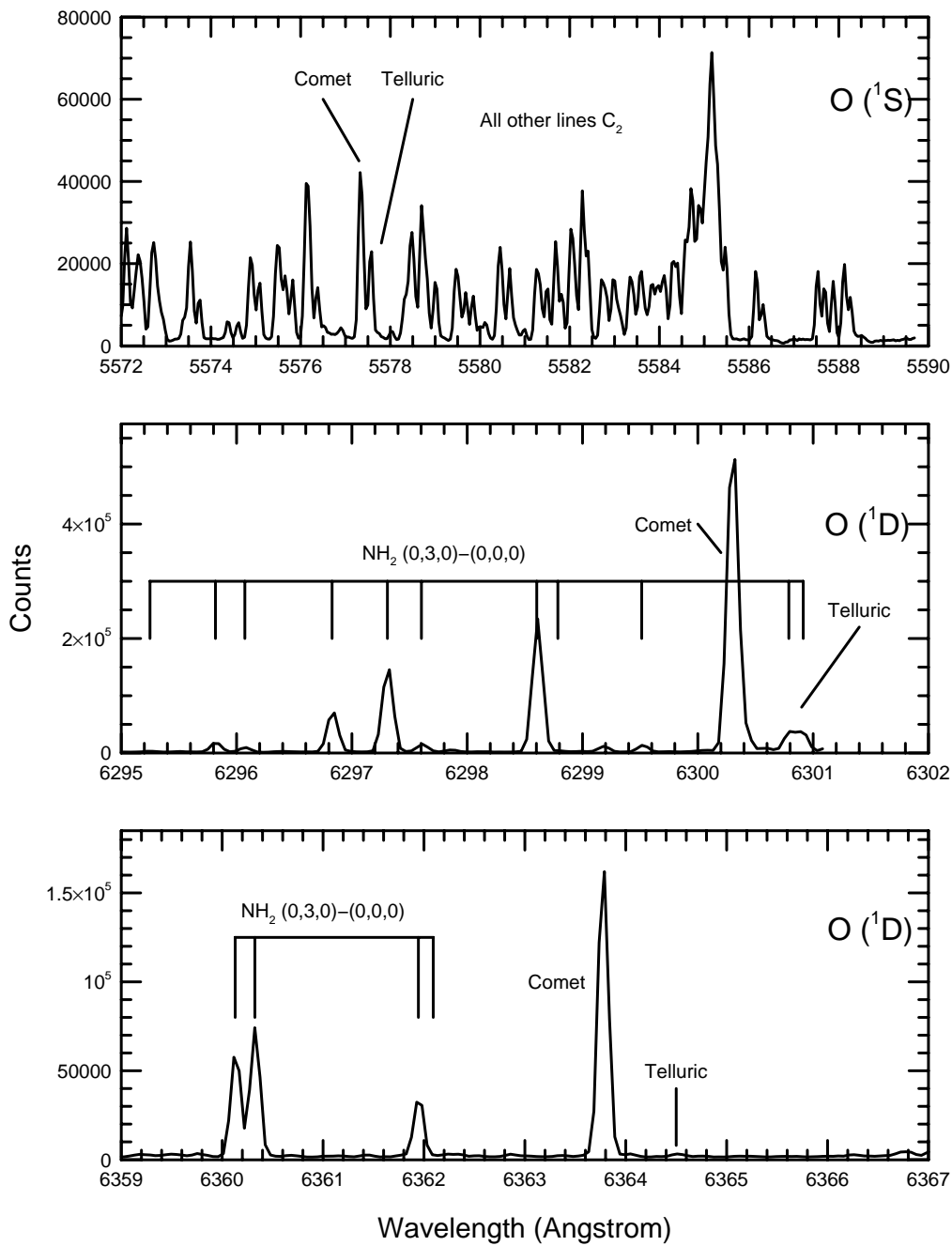}
\caption[fig2]{The spectral regions around
the three atomic oxygen lines for comet 122P/de~Vico.  These
spectra are typical of comets with moderately strong C$_{2}$ bands,
showing the potential C$_{2}$ contamination of the cometary green line.
Note also
the telluric oxygen line is Doppler-shifted from the cometary line.
The cometary oxygen line can be deblended from the contaminating
cometary C$_{2}$ and the telluric oxygen.  In contrast, the
spectral regions of the red lines are quite clean and only
the telluric line needs to be deblended.  }\label{devico}
\end{figure}

\begin{figure}
\vspace{7in}
\includegraphics{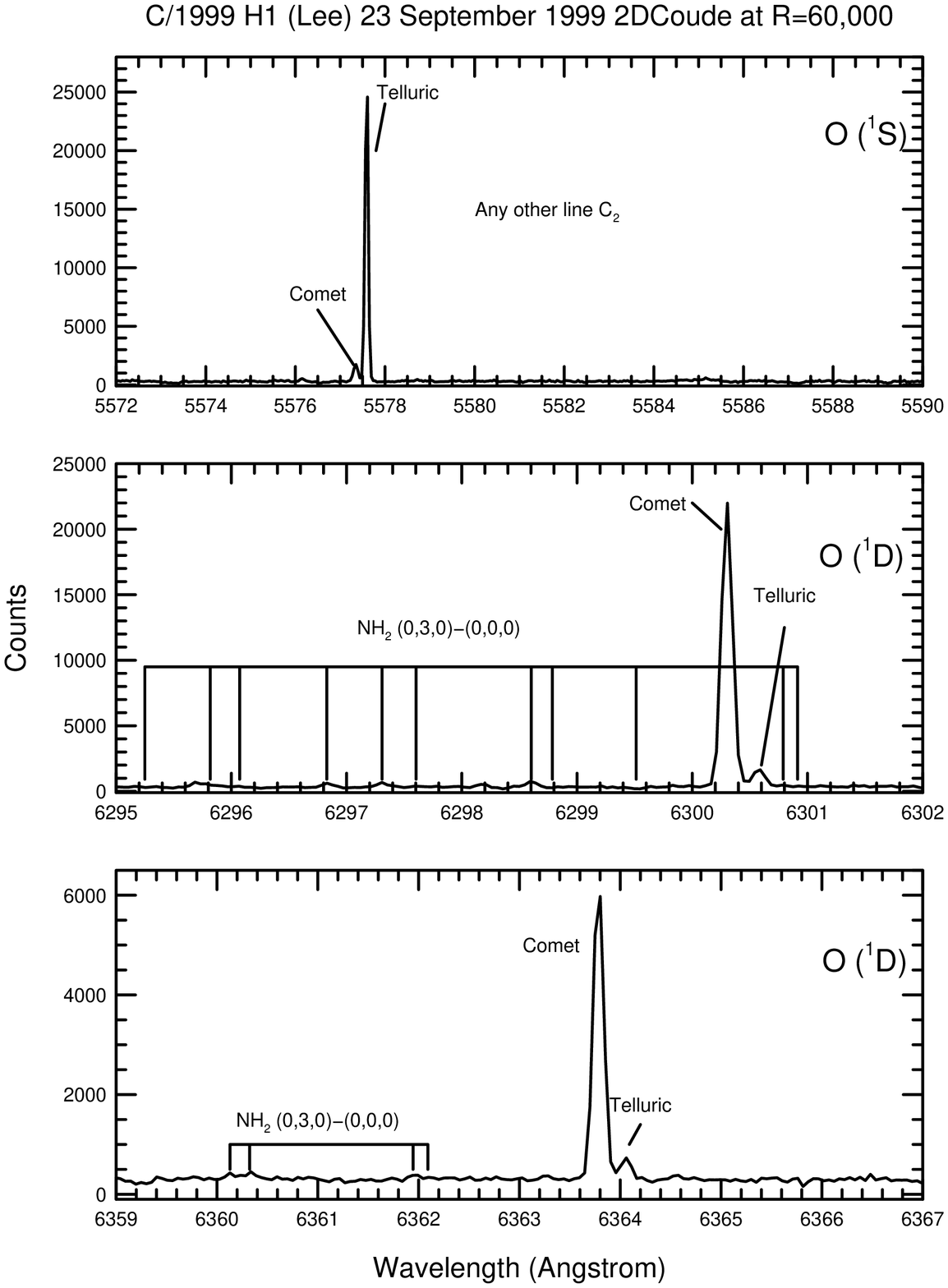}
\caption[fig3]{These spectra of comet C/1999 H1 (Lee)
show a case of a comet with relatively weak C$_{2}$.  It is easy
to see the cometary green line and the only contaminant is the
telluric line.  As with de~Vico, the red lines are in regions
that are not busy.  Again, the telluric and cometary lines can
be easily deblended.  }\label{lee}
\end{figure}

We observed 8 comets for which we were able to detect cleanly both the
green and red lines using the R=60,000 mode.  The circumstances
of the observations are listed in Table~\ref{log}. 
The slit was $1.2 \times 8.2$\,arcsec for all of these observations.
On some nights,
multiple spectra were obtained.  The number of spectra for each night
is listed in the table and includes only spectra centered on the
optocenter.
The reduction procedure followed was similar to that in
Cochran and Cochran (2001). \nocite{coco01}
Note that results for comet 1999 S4 (LINEAR), which were
presented in Cochran and Cochran, have been included in this
paper and were re-analyzed to ensure totally consistent processing.
The results are generally the same as in the earlier paper.
Figure~\ref{devico} shows a typical spectrum where the C$_{2}$ is
strong relative to the green oxygen line.  Figure~\ref{lee} is an
example of a comet with very little C$_{2}$ contamination of the
green line.
\begin{table}
\renewcommand{\baselinestretch}{1}
\caption{Observational Circumstances}\label{log}
\vspace*{5pt}
\centering
{\small
\begin{tabular}{lr@{ }l@{ }lccccc}             % 9 columns
\hline
\multicolumn{1}{c}{Comet} & \multicolumn{3}{c}{Date} & No. &
Heliocentric & Heliocentric & Geocentric & Geocentric \\
      & & & & Spectra & Distance & Radial & Distance & Radial \\
 & & &  & & & Velocity & & Velocity \\
      &  & & & & (AU) & (km s$^{-1}$) & (AU) & (km s$^{-1}$) \\
\hline
122P/de Vico & 3 & Oct & 1995 & 2 & 0.66 & -3.41 & 1.00 & -14.25 \\
             & 4 & Oct & 1995 & 1 & 0.66 & -2.29 & 0.99 & -12.87 \\
C/1996 B2 (Hyakutake) & 9 & Mar & 1996 & 2 & 1.37 & -32.80 & 0.55 & -57.30 \\
C/1999 H1 (Lee) & 30 & May & 1999 & 1 & 1.09 & -23.85 & 1.03 & 34.37 \\
                & 21 & Sep & 1999 & 2 & 1.51 & 24.97 & 0.87 & -16.48 \\
                & 23 & Sep & 1999 & 11 & 1.54 & 24.94 & 0.86 & -13.39 \\
D/1999 S4 (LINEAR) & 25 & Jun & 2000 & 4 & 0.97 & -19.70 & 1.21 & -61.21 \\
                   & 26 & Jun & 2000 & 5 & 0.96 & -19.38 & 1.17 & -61.61 \\
                   & 6 & Jul & 2000 & 4 & 0.86 & -15.08 & 0.81 & -62.84 \\
                   & 7 & Jul & 2000 & 4 & 0.85 & -14.53 & 0.77 & -62.46 \\
                   & 8 & Jul & 2000 & 2 & 0.84 & -13.96 & 0.74 & -61.92 \\
                   & 9 & Jul & 2000 & 1 & 0.84 & -13.36 & 0.70 & -61.19 \\
                   & 15 & Jul & 2000 & 1 & 0.80 & -9.28 & 0.50 & -50.11 \\
                   & 16 & Jul & 2000 & 2 & 0.79 & -8.52 & 0.45 & -46.46 \\
                   & 17 & Jul & 2000 & 5 & 0.79 & -7.74 & 0.45 & -42.03 \\
C/2001 A2 (LINEAR) & 24 & Jul & 2001 & 1 & 1.35 & 23.58 & 0.42 & 21.45 \\
C/2002 V1 (NEAT) & 19 & Jan & 2003 & 1 & 0.98 & -40.30 & 0.89 & 8.20 \\
C/2001 Q4 (NEAT) & 25 & May & 2004 & 5 & 0.98 & 4.92 & 0.66 & 45.87 \\
                 & 26 & May & 2004 & 4 & 0.98 & 5.45 & 0.68 & 46.13 \\
                 & 27 & May & 2004 & 3 & 0.98 & 5.96 & 0.71 & 46.32 \\
C/2006 M4 (Swan) & 31 & Oct & 2006 & 2 & 1.00 & 19.48 & 1.02 & 13.08 \\
\hline
\end{tabular}

}
\end{table}

We measured the intensity and the full width half maximum (FWHM) of a Gaussian
fit to the three oxygen lines 
for each of the spectra described in Table~\ref{log}.
When the telluric line was not
cleanly separated by the Doppler shift of the cometary
lines, we would simultaneously fit Gaussians
to both the cometary and telluric lines in order to deblend the
two features.  Each spectrum was measured independently and
an average was computed from all the observations for each comet for each line.

The measured spectral line width is a convolution of the intrinsic
width of the line (which is dominated by 
the velocity dispersion of the atoms from
their production) and the instrumental slit profile.
In order to define the instrumental line width, we used our
observations of a ThAr hollow cathode lamp.  The Th lines are intrinsically
quite narrow and are unresolved by our spectrograph.  We measured the
FWHM of 15 -- 30 isolated Th lines per spectral order of interest for each night
to define
the instrumental line profile for that order.  As would be expected, 
the instrumental
profile is slightly wider for the redder wavelengths under study.
We found very consistent instrumental line widths over the many years of
observations and the slight differences are probably the result
of differences in focus.  Typically, we could measure the 
FWHM of the instrumental profiles to 0.003~\AA.
The intrinsic cometary line width is then just the measured
line width with the instrumental width removed in quadrature.
The resultant intrinsic line widths can then be converted to 
a velocity width.

In addition to our R=60,000 resolution spectra listed in Table~\ref{log},
we also observed comet Hyakutake with the R=200,000 mode on 30 Mar 1996
($R_h$=0.94 AU, $\Delta$=0.19 AU).  With these observations we
only detected the green line and the 6300~\AA\ line, while the
6364~\AA\ line fell in an interorder gap.  In the R=200,000
mode, the intrinsic cometary line width is much wider than the instrumental
width.  

\section{Constraints From Our Spectra}
The intensity ($I$) of an emission line is dependent on the
photodissociative lifetime of the parent ($\tau_p$), the yield
of photodissociation ($\alpha$), the branching ratio for the
transition ($\beta$), and the column density of the parent ($N$) as
\[ I = 10^{-6} \tau_p^{-1} \alpha \beta N \hspace*{3em}{\rm photons\,cm^{-2}\,s^{-1}} \]

\begin{figure}
\vspace{4in}
\includegraphics{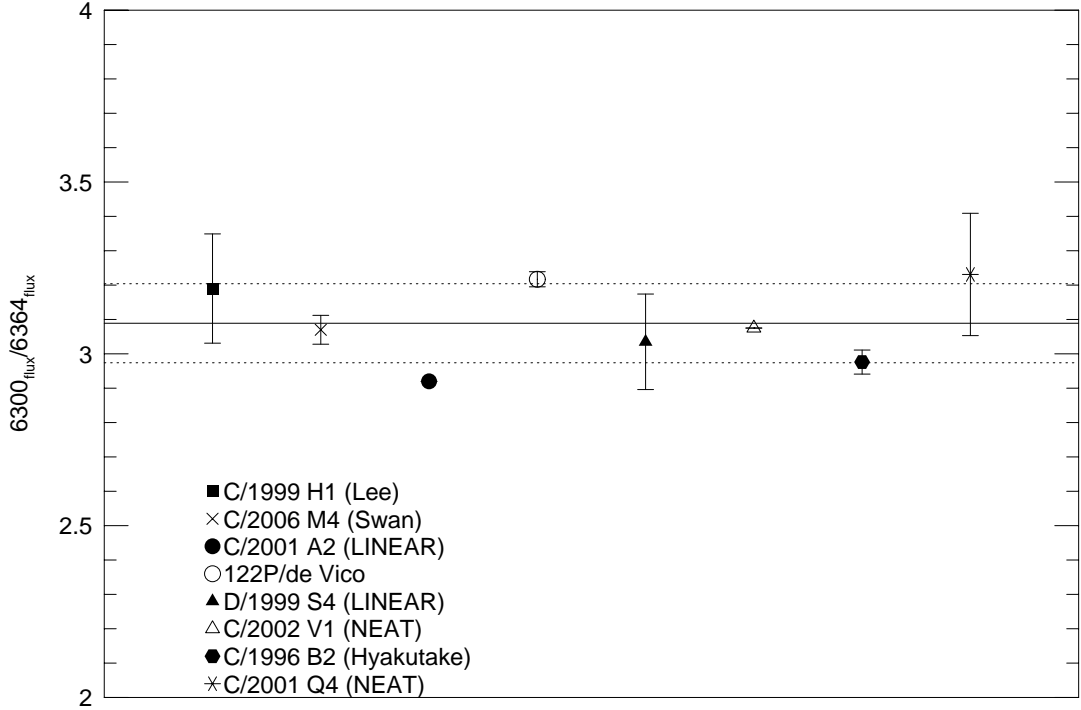}
\caption[fig4]{The average ratio of the two lines of
the red oxygen doublet for each of the eight comets is shown. The average
value of the eight comets is shown as a solid line and the 1$\sigma$
error range is denoted with dashed lines.  }\label{fluxred}
\end{figure}

The red lines are both transitions from the (2p$^4$)~$^1$D to the
(2p$^4$)~$^3$P ground state so their lifetimes, column density
and yields should be the same.  Thus, the ratio of their fluxes should
be the same as the ratio of their branching ratios.  Storey and Zeippen (2000)
have derived a theoretical value of 2.997 for the red doublet ratio based on
the Einstein A values.  Sharpee and Slanger (2006) measured this ratio
in terrestrial nightglow spectra and found a ratio of $2.997\pm0.016$.
\nocite{stze2000,shsl2006}  
The average value for each observed comet is listed in Table~\ref{flux}
and plotted in Fig.~\ref{fluxred}. 
We find an average value
of $3.09\pm0.12$ for these eight comets, which is in agreement
within the error bars with the theoretical and nightglow values.
\begin{table}
\renewcommand{\baselinestretch}{1}
{\small
\caption{Measured Line Ratios}\label{flux}
\vspace*{5pt}
\centering
\begin{tabular}{lcc}
\hline
\multicolumn{1}{c}{Comet} &
\multicolumn{1}{c}{6300/6364} &
\multicolumn{1}{c}{5577/(6300+6364)} \\
&
\multicolumn{1}{c}{flux ratio} &
\multicolumn{1}{c}{flux ratio} \\
\hline
122P/de Vico          & 3.22 $\pm$ 0.02 & 0.08 $\pm$ 0.003  \\
C/1996 B2 (Hyakutake) & 2.98 $\pm$ 0.04 & 0.09 $\pm$ 0.001  \\
C/1999 H1 (Lee)       & 3.19 $\pm$ 0.16 & 0.08 $\pm$ 0.01  \\
D/1999 S4 (LINEAR)    & 3.04 $\pm$ 0.14 & 0.06 $\pm$ 0.01  \\
C/2001 A2 (LINEAR)    & 2.92       & 0.11         \\
C/2002 V1 (NEAT)      & 3.08       & 0.09         \\
C/2001 Q4 (NEAT)      & 3.23 $\pm$ 0.18 & 0.09 $\pm$ 0.02  \\
C/2006 M4 (Swan)      & 3.07 $\pm$ 0.04 & 0.09 $\pm$ 0.01  \\
\hline
{\bf Average}      & 3.09 $\pm$ 0.12 & 0.09 $\pm$ 0.014 \\
\hline
\end{tabular}

}
\end{table}

\begin{figure}
\vspace{4in}
\includegraphics{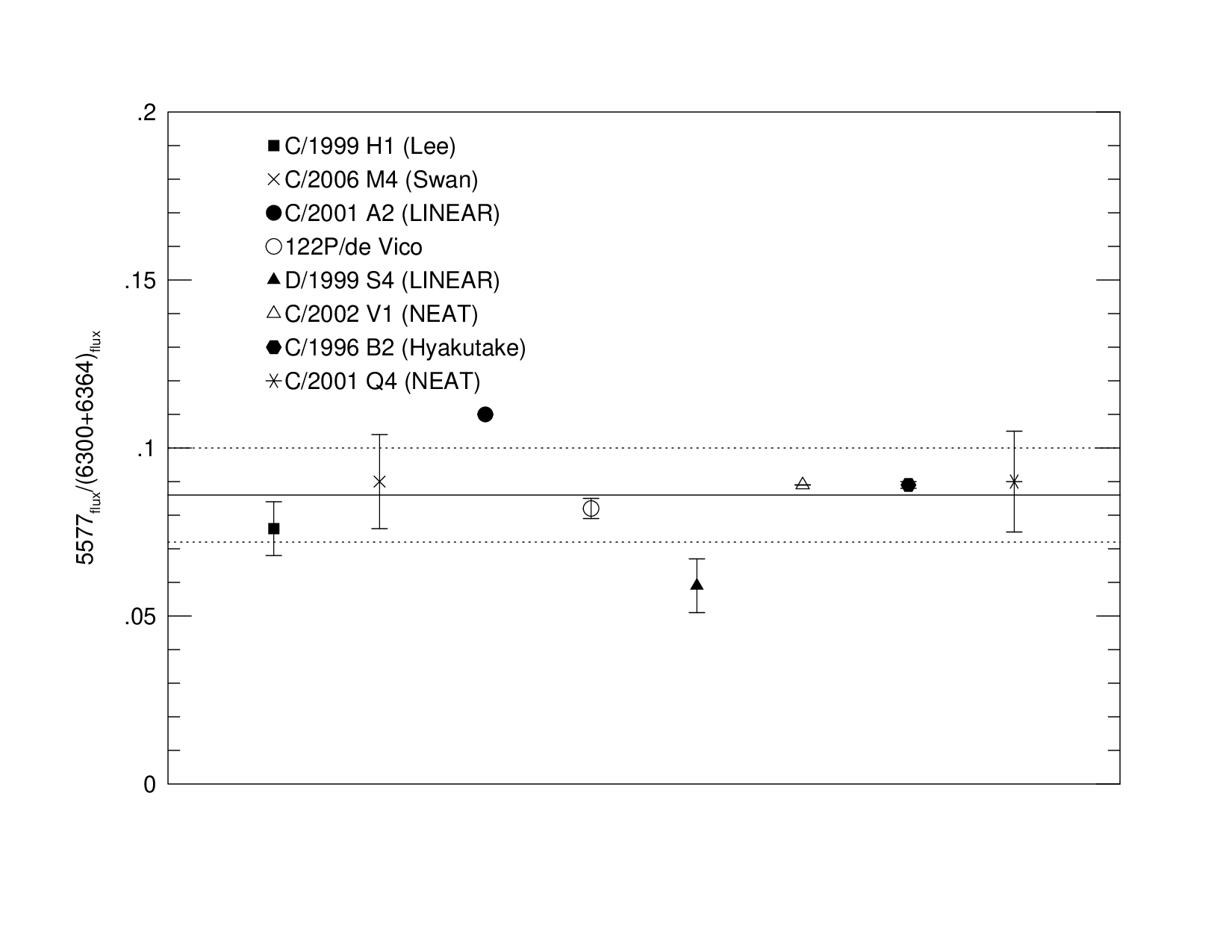}
\caption[fig5]{The average ratio of the green
line to the sum of the two lines of
the red oxygen doublet for each of the eight comets is shown. The average
value of the eight comets is shown as a solid line and the 1$\sigma$
error range is denoted with dashed lines.  }\label{fluxgreenred}
\end{figure}

In the absence of collisional quenching, the
ratio of the O~($^1$S -- $^1$D) intensity to the sum of the
O~($^1$D -- $^3$P) intensities can be expressed as
\[ \frac{I_{5577}}{(I_{6300} + I_{6364})} =
   \frac{\tau^{-1}_{p(^1S)} \alpha_{(^1S)} \beta_{5577} N_{(^{1}S)} }
       {\tau^{-1}_{p(^1D)} \alpha_{(^1D)} (\beta_{6300} + \beta_{6364}) N_{(^{1}D)} }
\]
In order to evaluate this equation, one needs to know the parent(s) of the
oxygen to know what values of lifetimes, yields and branching ratios
to use. 
Festou and Feldman (1981) \nocite{fefe81} listed in their
Table~3 some effective excitation rates for the dissociation.
These are reproduced in Table~\ref{dissociate}.
Festou and Feldman estimate that these values are only accurate
to $\pm50$\% because of variations in the solar EUV flux.
Inspection of this table shows that we expect a very different
value for the green-to-red line ratio for H$_{2}$O than for CO and
CO$_2$.  Thus, the observed values for this line ratio may shed
light on the potential parent.  The ratios we find are listed
in Table~\ref{flux}. 
These ratios have been corrected for the 
relative response of the red and green orders of the spectrograph
by use of observations of $\alpha$~Lyr (the green-to-red flux ratio needed to
be increased by 30\%, as per Cochran and Cochran 2001). 
Figure~\ref{fluxgreenred} shows
our measured values for the ratio of the green line flux to
the sum of the line fluxes from the two red lines.  We find an average
value of $I_{5577}/(I_{6300} + I_{6364}) = 0.09\pm 0.01$.
Using the values of Festou and Feldman,
this is consistent with water as the parent of the oxygen.
\begin{table}
\renewcommand{\baselinestretch}{1}
{\small
\caption{Festou and Feldman (1981) Effective Excitation Rates}\label{dissociate}
\vspace*{5pt}
\centering
\begin{tabular}{lrrc}
\hline
 & \multicolumn{2}{c}{Excitation Rate} & Ratio \\
 & \multicolumn{2}{c}{(s$^{-1}$ at 1AU)} & \\
\cline{2-3}
\multicolumn{1}{c}{Parent} & O ($^1$S) & O ($^1$D) & O($^1$S)/O($^1$D) \\
\hline
H$_{2}$O & 7--12$\times10^{-8}$ & 8--12$\times10^{-7}$ & $\sim0.1$ \\
CO & $<4\times10^{-8}$ & $<4\times10^{-8}$ & $\sim1$ \\
CO$_2$ & 4.4$\times10^{-7}$ & 5$\times10^{-7}$ & $\sim1$ \\
\hline
\end{tabular}

}
\end{table}

The slit covered from $\sim1200$ to $\sim3500$~km from the optocenter
and thus included the highest density regions of the coma.  It is only
in this very inner coma that collisional quenching would have any effect.
The collisional quenching coefficients of O~($^1$S) and O~($^1$D) by
H$_{2}$O are comparable but the O~($^1$D) lifetime is $\sim150$ times
longer than the O~($^1$S) lifetime.  Festou and Feldman (1981) showed that
this will result in a flatter column density profile for the red
lines than for the green line.  In addition, the red line intensities
might be lower than they would be in the absence of quenching.

If the parent is H$_{2}$O then the reactions which might apply are
\begin{eqnarray*}
\mbox{H$_{2}$O} + h\nu & \rightarrow & \mbox{H} + \mbox{OH} \\
\mbox{H$_{2}$O} + h\nu & \rightarrow & \mbox{H$_2$} + \mbox{O ($^1$S)} \\
\mbox{H$_{2}$O} + h\nu & \rightarrow & \mbox{H$_2$} + \mbox{O ($^1$D)} \\
\mbox{H$_{2}$O} + h\nu & \rightarrow & \mbox{2H} + \mbox{O ($^3$P)} \\ [-10pt]
\end{eqnarray*}
Huestis (2006) points out that production of O ($^1$S) by photodissociation
of H$_{2}$O has never been reported in the laboratory literature.  He goes
on to claim that no experiment has been attempted so that there is
no experimental basis for the estimate of its yield.
\nocite{hu2006}

As noted in the introduction, the further decay of OH into some form
of oxygen was shown not to be an important source of O ($^1$S) because
the decay of OH flux after the impact of the Deep Impact spacecraft with
Tempel 1 was much slower than that for the green line flux.

If the parent is CO$_2$ then the relevant reactions are
\begin{eqnarray*}
\mbox{CO$_2$} + h\nu & \rightarrow & \mbox{CO} + \mbox{O ($^1$D)} \\
\mbox{CO$_2$} + h\nu & \rightarrow & \mbox{CO} + \mbox{O ($^1$S)} \\
\mbox{CO$_2$} + h\nu & \rightarrow & \mbox{CO} + \mbox{O ($^3$P)} \\ [-10pt]
\end{eqnarray*}
Huestis (2006) points out that for these equations the yield of O ($^1$S) has
been measured in a number of studies and it approaches unity
in a narrow window around 112.5\,nm.  However, for CO$_2$, the
yield of O ($^1$D) has never been measured.  This experiment is
difficult because of the rapid quenching,  but it is believed
that O ($^1$D) is the primary product over much of the absorption
spectrum.

The production of oxygen from the dissociation of CO is possible, but it is 
slower than the production from H$_{2}$O and CO$_2$ by up to an
order of magnitude because of the bond strength of the CO molecule
(Bhardwaj and Haider 2002).  \nocite{bhha2002}
In addition, Cochran and Cochran (2001)
showed that the production of the oxygen from CO in comet
1999~S4 (LINEAR) was inconsistent with UV observations showing
a depletion of CO in its coma.
Thus, we dismiss CO as a parent. 

In addition to measuring the line intensities for the three oxygen lines,
we also measured the line widths listed in Table~\ref{widths}.  
The derived intrinsic line widths, after removal of the instrumental
profile, in both \AA\ and km s$^{-1}$, are tabulated.
\begin{table}
\renewcommand{\baselinestretch}{1}
{\small
\caption{Measured Oxygen Line Widths}\label{widths}
\vspace*{5pt}
\centering
\begin{tabular}{lcccccc}
\hline
\multicolumn{1}{c}{Comet} &
\multicolumn{2}{c}{5577 Intrinsic} &
\multicolumn{2}{c}{6300 Intrinsic} &
\multicolumn{2}{c}{6364 Intrinsic} \\
 & FWHM & velocity & FWHM & velocity & FWHM & velocity \\
 & \AA & km s$^{-1}$ & \AA & km s$^{-1}$ & \AA & km s$^{-1}$ \\
\hline
122P/de Vico          & 0.073$\pm$0.031 & 2.4 & 0.034$\pm$0.016 & 1.0 & 0.025$\pm$0.018 & 0.7  \\
C/1996 B2 (Hyakutake) & 0.080$\pm$0.011 & 2.6 & 0.073$\pm$0.009 & 2.1 & 0.043$\pm$0.015 & 1.2  \\
C/1999 H1 (Lee)       & 0.085$\pm$0.038 & 2.7 & 0.036$\pm$0.028 & 0.6 & 0.049$\pm$0.026 & 0.8  \\
D/1999 S4 (LINEAR)    & 0.067$\pm$0.011 & 2.2 & 0.050$\pm$0.008 & 1.4 & 0.058$\pm$0.012 & 1.6  \\
C/2001 A2 (LINEAR)    & 0.058$\pm$0.011 & 1.9 & 0.009$\pm$0.009 & 0.3 & 0.040$\pm$0.044 & 1.1  \\
C/2002 V1 (NEAT)      & 0.095$\pm$0.011 & 3.1 & 0.046$\pm$0.038 & 1.3 & 0.052$\pm$0.042 & 1.5  \\
C/2001 Q4 (NEAT)      & 0.079$\pm$0.007 & 2.5 & 0.031$\pm$0.017 & 0.9 & 0.033$\pm$0.018 & 0.9  \\
C/2006 M4 (Swan)      & 0.080$\pm$0.038 & 2.6 & 0.049$\pm$0.014 & 1.4 & 0.057$\pm$0.011 & 1.6  \\
\hline
\multicolumn{7}{p{5in}}{Errors are the combined standard deviations from the multiple measurements of each comet and the multiple lines of Th except for 2001 A2 and 2001 Q4.  There was only a single spectrum for each of these comets
so the instrumental measurement error was used as the comet error.}
\end{tabular}

}
\end{table}

\begin{figure}
\vspace{4in}
\includegraphics{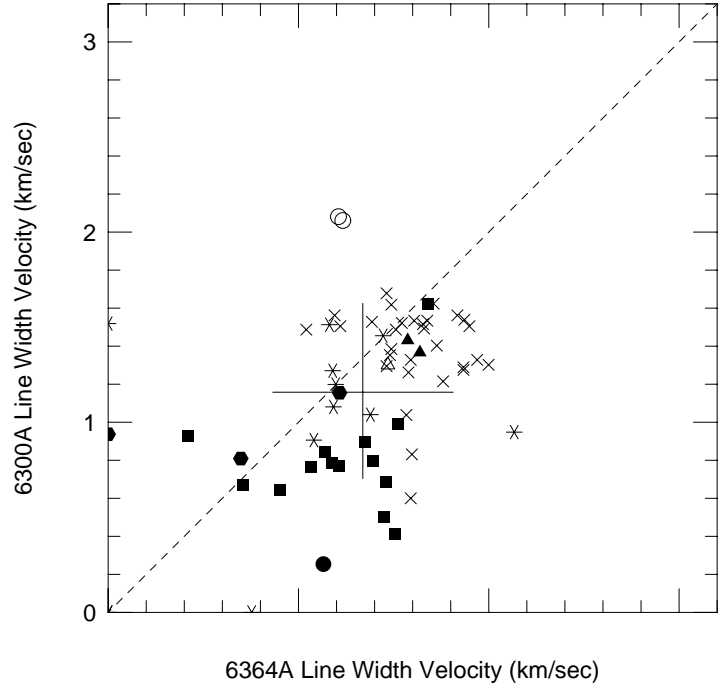}
\caption[fig6]{The line widths from the individual
spectra of the eight comets are plotted for the two lines of the
red doublet.  The comets are denoted with the
same symbols as in Figures~\ref{fluxred} and \ref{fluxgreenred}.
The diagonal dotted line is the equality line.  The large plus marks
the average values for each width and their standard deviations. }\label{redwidths}
\end{figure}

The red lines can be formed
without forming the green line. Since both transitions are from
the same upper state to the ground state, they should have the same line
width. Figure~\ref{redwidths} shows the 6300~\AA\ line velocity plotted
against the 6364~\AA\ line velocity for the individual spectra on each
comet.  In a few cases, the uncertainties in the line widths made
it impossible to discern any width greater than the instrumental
width.  These points are plotted with a width of 0\,km s$^{-1}$ in this
figure. With the exception of these few spectra with lines which could not
be measured, the two lines of the red doublet have the same width.

\begin{figure}
\vspace{4in}
\includegraphics{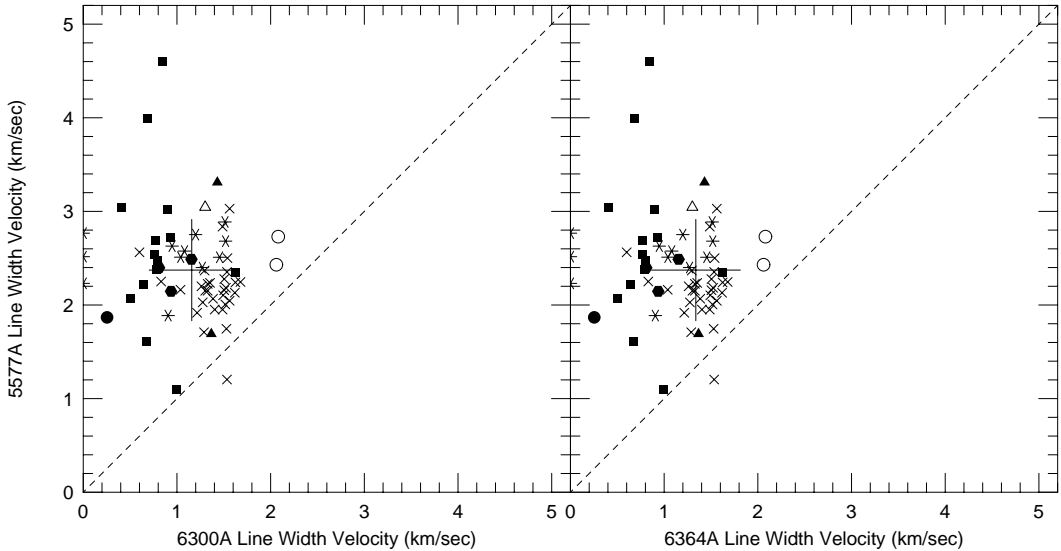}
\caption[fig7]{The line widths for
the green lines of the individual spectra are plotted against the
widths of the 6300~\AA\ line in the left side of the plot and
for 6364~\AA\ in the right side.  The dotted line denotes equality.
The average values and standard deviations are shown with the
big plusses.  The filled symbols are for comets, such as Lee,
where the O ($^1$S) line dominates the C$_{2}$ features; the open
symbols are for comets such as de~Vico where we had significant
C$_{2}$ signal.  }\label{greenredwidths}
\end{figure}

If the green line
is present, it will decay to the red doublet 90--95\% of the time.
Thus, if the red and green lines are produced by the same process and the same
energy photons, then
they should be the same widths.  Figure~\ref{greenredwidths} shows
the width of the green line in comparison with each of the red lines.
In these observations, the green line is {\it not} the same 
width as the red lines.  
\nocite{fe81b}
Festou (1981) listed the excess velocity for the creation of the O ($^1$S)
state from Ly $\alpha$ photons as 1.6\,km\,s$^{-1}$ and of the O ($^1$D)
state as 1.8 km\,s$^{-1}$. Thus, there would be the expectation that
the red lines were broader than the green line.
However, this is not what we observe. 
Figure~\ref{greenredwidths} shows
that it is the green line which is the widest of the lines.
Inspection of the data also shows that we get the same result
for comets with very weak C$_{2}$ as we do for those with stronger
C$_{2}$, so our deblending was quite successful.
The difference in the line widths for the green and red lines
must point to additional parent(s) for the green line or to dissociation
to the $^1$S state by photons with higher energy than Ly~$\alpha$, the
presumed exciter of the $^1$D state.

\begin{figure}
\vspace{7in}
\includegraphics{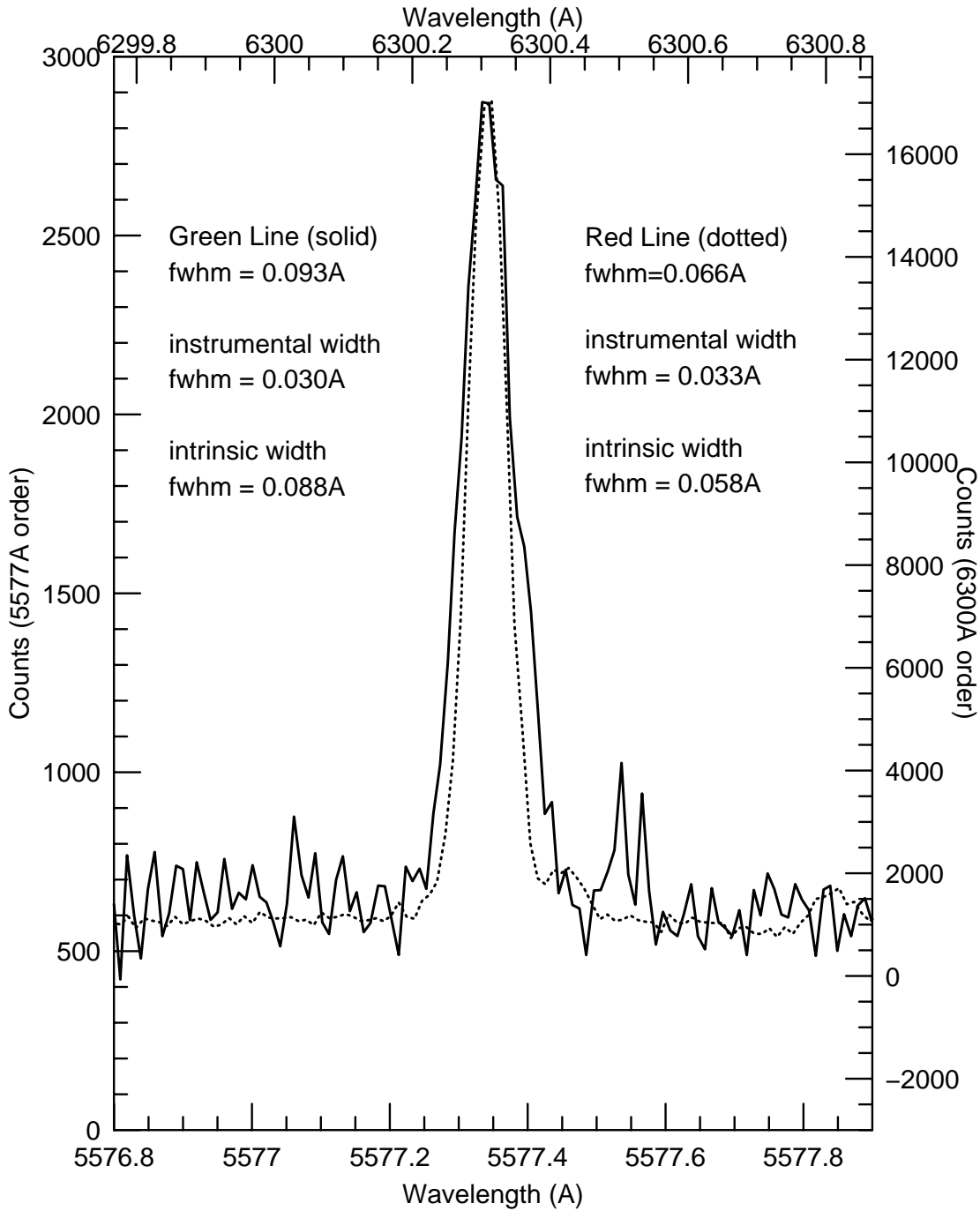}
\caption[fig8]{The green line (solid; labels left
and bottom) and the
6300~\AA\ line (dotted; labels right and top) from the 30 Mar 1996 R=200,000
spectrum of Hyakutake are plotted together.  The spectra are normalized to the
same relative intensity and the wavelength scales are the same
number of \AA/inch.   The profiles are the convolution
of the instrumental and intrinsic line widths. The instrumental width
is wider for the red line but the observed green line is much wider.
Thus, the green line must be intrinsically a wider line.  }\label{hyakutake}
\end{figure}

In addition to the R=60,000 observations shown in Table~\ref{widths},
we have three spectra of comet Hyakutake
from 30 March 1996 obtained with the R=200,000 mode.  In this
setup, we cannot observe all three lines simultaneously, but we
were able to observe the 5577~\AA\ and 6300~\AA\ lines in
one spectral image.  Figure~\ref{hyakutake} is a plot of
the green line and the 6300~\AA\ line from these observations.
At this resolving power, the instrumental widths are a much
smaller fraction of the cometary line widths than
with R=60,000.  As with the R=60,000
spectra, the instrumental width of the red line is wider than that
of the green line, but the intrinsic cometary line width of
the green line is wider than that of the red line.  
Thus, our conclusions from Figure~\ref{greenredwidths} are not biased by
the lines being just a bit wider than the instrumental widths
at the lower resolving power.

\begin{figure}[ht!]
\vspace{4.5in}
\includegraphics{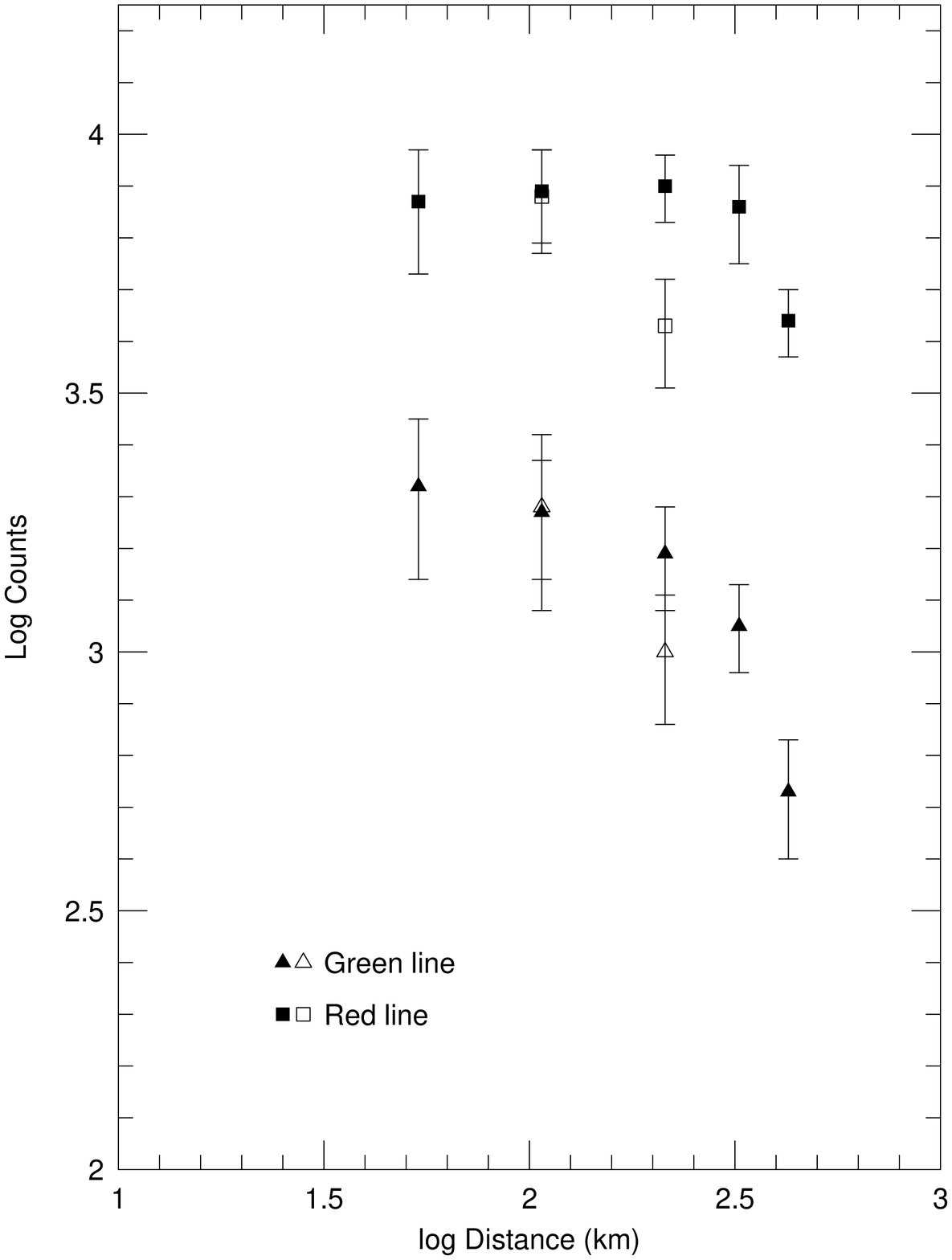}
\caption[fig9]{ The flux in the green and red line
was extracted in seven pieces along the slit from the R=200,000 spectra
of comet Hyakutake
and is plotted against cometocentric distance.
The closed symbols are the data on one side of the optocenter
and the open symbols the other (the comet was not centered in the slit).
The open data points at 215\,km appear low, possibly due to vignetting
at the end of the slit.  The red line data shows a much flatter trend
with cometocentric distance than does the green line.  }\label{hyatrend}
\end{figure}

Comet Hyakutake was quite close to the Earth ($\Delta = 0.19$\,{\sc au})
when we observed it at R=200,000.  Thus, each pixel in the spatial
direction along our 8 arcsec slit subtended only 17.9\,km at the comet.
We extracted the spectra in 7 chunks along the slit to investigate
whether the line width changed in the very inner coma.  We found
that the intrinsic FWHM was constant across the slit for both
the green and red lines, with the average FWHM of the green line
being $0.088\pm0.003$~\AA\ and the red line being $0.056\pm0.002$~\AA.
Thus, the line widths from extraction along the whole slit, shown
in Fig.~\ref{hyakutake}, are not widened by outflow in the
inner 400\,km of the coma.  In addition, the line intensity of the
green line decreased more quickly with cometocentric distance
than did the red line.
This is shown in Figure~\ref{hyatrend}.  The flat distribution
of the red line flux out to 400 km (and possibly the green line flux
out to 100 km) is consistent with quenching in the inner coma.  These
profiles agree well with Case B of Festou and Feldman (1981; their figs.
2 and 3).

\section{Discussion}
Spectra of comets tell us about the physical and chemical processes
taking place within the cometary comae.  In addition, comets are
excellent laboratories for testing our understanding of chemistry
since the long path lengths through the coma allow relatively
small gas densities to produce significant column densities.

We are interested in the oxygen lines for several reasons.  First,
oxygen is one of the most abundant elements in the Universe
after hydrogen and helium and it readily bonds with other atoms
and molecules.  Indeed, 80\% of cometary ices are H$_{2}$O, so
oxygen is a critical component of comets.  Water is hard to
detect spectrally from the ground so daughter products of water
have traditionally been used as a proxy to study the water.
Since the OH fluorescent bands are in the UV region of the spectrum,
where there is low detector quantum efficiency and high atmospheric
extinction, they 
are difficult to study.  The three forbidden oxygen lines
discussed in this paper are in the optical and
have been used as a water proxy.

Unfortunately for studies of atomic oxygen in comets, the telluric
spectrum has strong atomic oxygen emissions which are always present
and are quite variable.  Thus, until the advent of the era of
high resolution spectroscopy of comets, contamination
of the cometary lines by the telluric lines was always an issue.
Even with high resolving powers, the green line in cometary
spectra has proved to be a difficult target since it sits
within the spectral region of the C$_{2}$ (1,2) P-branch.
Despite that, there have been some observations 
of the detection of the green line in cometary spectra previously reported.
These are listed in Table~\ref{otherdata}.
\begin{table}
\renewcommand{\baselinestretch}{1}
{\small
\caption{Previously Reported Green Line Detections}\label{otherdata}
\vspace*{5pt}
\centering
\begin{tabular}{lccccl}
\hline
\multicolumn{1}{c}{Comet} & Date(s) & Heliocentric & Geocentric &
$\frac{I_{5577}}{(I_{6300}+I_{6364})} $ &
Ref.\\
 & & Distance & Distance & flux & \\
 & & (AU) & (AU) & ratio & \\
\hline
C/1983 H1 & 6, 8 May 1983 & 1.02 & 0.13, 0.09 & 0.022--0.034 & 1 \\
\hspace*{2em}(IRAS-Araki-Alcock) & \\
1P/Halley & 5 Apr 1986 & 1.25 & 0.45 & 0.05 -- 0.1 & 2 \\
C/1996 B2 (Hyakutake) & 23, 27 Mar 1996 &  1.08, 1.00 & 0.12 &  0.13 -- 0.16 & 3 \\
C/1995 O1 (Hale-Bopp) & 26, 28 Mar, 22 Apr 1997 & 0.92, 0.99 & 1.33, 1.63 & 0.18 -- 0.22 & 4 \\
153P/Ikeya-Zhang & 20 Apr 2002 & 0.89 & 0.43 & 0.12$\pm$0.1 & 5 \\
9P/Tempel 1 & 2 -- 6,9 Jul 2005 & 0.89--0.92 & 1.51 & 0.056-0.13$\pm$0.1 & 6 \\
\hline
\multicolumn{6}{p{5.5in}}{References -- 1: Cochran (1984) - red and green lines observed different
nights; 2: Smith and Schempp (1989) - 6364 line not observed; 3: Morrison
{\it et al.} (1997) - 6300 line not observed; 4: Zhang {\it et al.} (2001);
5: Capria {\it et al.} (2005); Capria {\it et al.} (2008)} \\
\hline
\nocite{co84,smsc89,moetal97,zhzhhu01,caetal2005,caetal2008}
\end{tabular}

}
\end{table}

The three oldest observations all suffered from non-simultaneity of the
observations of the three lines.  Indeed, the Halley and Hyakutake
observations reported in Table~\ref{otherdata} each contain
observations of only 1 line of the red doublet.  Thus, these
observations should be used cautiously.  The ratios of the green
line to the sum of the red line fluxes for Hale-Bopp, Ikeya-Zhang
and Tempel 1 are consistent with the values we found for our
8 comets reported in this paper.  Hale-Bopp seems to have the highest
ratio.  This might be because Hale-Bopp had extremely high water
production rates and a very large collisional zone.  However,
there are not a great deal of details in the Zhang {\it et al.} (2001) paper
so it is difficult to judge why this ratio might be higher. 
Hale-Bopp had a very strong continuum and C$_{2}$ band  so there may have been
issues correcting for these factors.

Comparing these remaining three comets with the 8 observations
presented here, it is obvious that the green line is typically about
10\% the summed intensity of the red doublet. 
If there is collisional quenching, as implied in Fig.~\ref{hyatrend},
the red line fluxes might be lower than without quenching so the 
$^1$S/$^1$D ratio in the absence of quenching might be even lower.
All of the authors
referenced in Table~\ref{otherdata}, and indeed our previous
paper on 1999 S4 (LINEAR) (Cochran and Cochran 2001), have
used this ratio along with Table~3 of Festou and Feldman (1981)
to conclude that water is the source of the green atomic oxygen line.

In this paper, we reported on the widths of the oxygen lines in addition
to their intensities and found that the green line was consistently
wider than the red lines.  Since 90--95\% of the photons in the
$^1$S state decay to the ground state via the $^1$D state, the higher
velocity photons should widen the red lines.  
We examined how this would
affect our measurements by generating two Gaussian line profiles
with FWHM values appropriate for the green line and assuming
the red line FWHM was the ``correct" value.   These two
lines were then combined with the same average flux ratio observed in 
the comets reported in this paper.  We then measured the properties of this
artificially generated line against a red line generated by
assuming none of the red line comes from photons which were first excited
to the $^1$S state.
We find the combined line has 2\% lower flux than a pure line
but that the FWHM of the lines agree to 0.001~\AA, or at least
as good as our measurement accuracy.  Therefore, based on the relative
intensity of the green and red lines, we would not have been able
to measure any broadening of the red lines and the measured red
line widths can be assumed to be indicative of excitation in
the $^1$D state only.

The wider green line leads to two possibilities for the production of
photons in the $^1$S state.  The first is that there is a parent other than
H$_{2}$O that dissociates into photons in the $^1$S state.  As discussed
above, CO can be eliminated based on its bonding and on observations
of comet 1999~S4.  OH can be eliminated because it decays more slowly
than the green line.  Thus, CO$_2$ would be the likely alternative
parent. 

The other possibility is that H$_{2}$O is the parent for both
the $^1$S and $^1$D states.  This would imply that the energy of the 
photons responsible for the production of the $^1$S state is higher than
that for the production of the $^1$D state.
As Huestis (2006) pointed out, these
excitation rates are not calibrated.  The laboratory measurement of the
production of the green line from H$_{2}$O has never been made.
Nor has the production of the red lines from CO$_2$.  Until
the point that these rates are known with any certainty, all we
can definitely conclude is that the green line flux is about 10\% that
of the combined red line flux in all comets.

However, this in itself may offer a clue.
The observations of the 8 comets reported here span 11 years, or a full solar
activity cycle.  They also span heliocentric distances from 0.66 to
1.54\,{\sc au}.  We see no difference in either the relative line
intensities
nor in the line widths from one comet to another.  Thus, whatever
photodissociative process(es) produce these lines, their relative 
importance does not change with solar UV flux or from comet to comet.
Inspection of Table~I of Bockel\'{e}e-Morvan {\it et al.} (2004)
\nocite{bocrmuwe2004}
shows that the abundances of minor species such as CO and CO$_2$ relative
to water are quite variable.  The constancy of the green line to the
red line flux thus suggests a single parent for both states with
different energies responsible for the production of the different states.
Clearly, more laboratory work is needed but comets are lending
clues to the processes involved.

\vspace*{0.75in}
\begin{center}Acknowledgments\end{center}
This paper includes data taken at The McDonald Observatory of The University of Texas at Austin.  This work was supported by NASA Grant NNG04G162G.
I thank D. Lambert, M. Endl, and W. Cochran for obtaining some
of the spectra.

\end{document}